\documentclass[doublecol]{epl2} 
\usepackage{amsmath,graphicx}

\title{Demographic noise can lead to the\\ spontaneous formation of species}

\author{T. Rogers\inst{1} \and A. J. McKane\inst{1}, \and A. G. Rossberg\inst{2}}
\shortauthor{T. Rogers \etal}
\institute{                    
  \inst{1} Theoretical Physics Division, School of Physics \& Astronomy, The University of Manchester, M13 9PL, UK\\
  \inst{2} School of Biological Sciences, Queen's University Belfast, BT9 7BL, UK
}
\pacs{05.40.-a}{Fluctuation phenomena, random processes, noise, and Brownian motion}
\pacs{87.23.Cc}{Population dynamics and ecological pattern formation}
\pacs{87.10.Ca}{Analytical theories}

\abstract{
When a collection of phenotypically diverse organisms compete with each other for limited resources, the population can evolve into tightly localised
clusters. Past studies have neglected the effects of demographic noise and studied the population on a macroscopic scale, where cluster formation is
found to depend on the shape of the curve describing the decline of competition strength with phenotypic distance. Here we show how including the
effects of demographic noise leads to a radically different conclusion. Two situations are identified: a weak-noise regime in which the population
exhibits patterns of fluctuation around the macroscopic description, and a strong-noise regime where clusters appear spontaneously even in the case that all organisms have equal fitness.}

\begin{document}

\maketitle
\section{Introduction}
Competitive interaction between organisms has long been recognised as a key component in the formation of species and, more widely, entire ecosystems.
A mathematical formulation of this idea was provided by MacArthur and Levins \cite{MacArthur1967} over forty years ago and variations of their model have been studied ever since, with considerable recent interest in the mechanisms by which an initially diverse population can aggregate into tightly clustered groups \cite{Sasaki1997,Fuentes2003,HernandezGarcia2004,Scheffer2006,Pigolotti2007,Pigolotti2009,Fort2010}. In these models, the phenotype of an organism is summarised by a single number $x$, representing a point in the `niche space' of all possible phenotypes. Writing $\Phi(x,t)$ for the size of the population at point $x$ in niche space and $t$ in time, a common approach \cite{Sasaki1997,Fuentes2003,Pigolotti2007} is to study a dynamical system of the form:
\begin{equation}
\begin{split}
\frac{\partial}{\partial t}\Phi(x,t)=&\,\Phi(x,t)+D\frac{\partial^2}{\partial x^2}\Phi(x,t)\\
&\quad-\frac{1}{K}\int \Phi(x,t)\Phi(y,t)g(x-y)\,dy\,.\quad
\end{split}
\label{macro}
\end{equation}
The three terms on the right of this equation correspond to (i) reproduction, where time has been scaled to give this rate one everywhere in niche space (ii) mutation, realised as diffusion in niche space with strength $D$, and (iii) death, given by the total effect of competition with all other organisms. The competition strength between organisms at points $x$ and $y$ in niche space is given by the competition kernel $g(x-y)$, while the parameter $K$ controls the overall carrying capacity of the ecosystem.\par
The homogeneous state with total population size $K$ is a fixed point of equation \eqref{macro}. Starting from this observation, a typical analysis considers the effect of small perturbations to the homogeneous state \cite{Pigolotti2007}. Often the perturbations die out and the population remains evenly spread, but in some cases fluctuations of a certain wavelength (in niche space) are amplified, eventually leading to the formation of sharp peaks separated by empty regions. Ecologically, these clusters of phenotypically similar organisms can be thought of as a prototypical `species', and the model can aid our understanding of the effect of competition on species formation. From a physical perspective, the emergence of clusters is a pattern-forming transition, and a mathematical analysis reveals that the crucial factor is the precise shape of the competition kernel; species can only form if $g$ has a negative Fourier mode and $D$ is sufficiently small. 
\par
One natural choice for the competition kernel is specified by the overlap between the organisms' consumption of resources, however, it is known that in this case all Fourier modes will always be positive, and thus the homogeneous state is stable \cite{Roughgarden1979}. In general, it may be difficult to find a simple ecological argument which justifies the choice of a pattern-forming kernel. Moreover, the common choice of a Gaussian kernel is a marginally stable case, meaning that a small change to the kernel can lead to very different behaviour. This lack of robustness of the pattern-formation mechanism has caused some debate in the literature \cite{Pigolotti2007,Pigolotti2009,Fort2010}. \par
Equation \eqref{macro} is intended to provide a macroscopic summary of the collective behaviour of a large number of individual organisms. In reality, the discreteness of the population means that it will experience some intrinsic demographic noise; a central assumption of the macroscopic theory is that this noise is not important. However, recent work has revealed that many complex systems experience significant noise driven phenomena at scales relevant to real world observations \cite{McKane2005,Butler2008}. In this article we present the results of a mathematical analysis of noise effects in phenotypic competition, with dramatic implications for the theoretical understanding of the interplay between competition and mutation in ecological systems.    \par
Our main result is the discovery of two distinct regimes of noise-induced phenomena, illustrated in Figure 1. This is achieved through the study of an individual-based stochastic model of phenotypic competition. We are able to show how the macroscopic theory \eqref{macro} may be derived directly from the stochastic model (rather than postulated {\it a priori}), but also how the effects of demographic noise can radically alter the observed behaviour. When the strength of competition is small in comparison with the rate of mutation, the model exhibits patterns of fluctuation around the homogeneous state. A more extreme effect is found when competition strength and mutation rate scale together. In this case the population can spontaneously cluster into stable macroscopic species, even in the case of a completely flat competition kernel, where no selective pressure is present at all. 

\begin{figure}
\includegraphics[width=0.45\textwidth, trim=10 0 45 0]{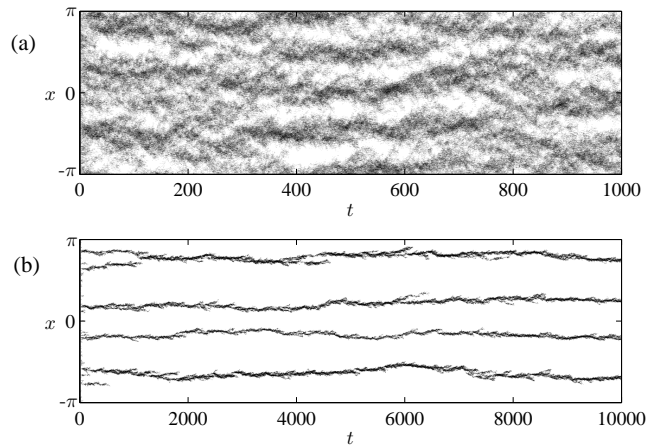}
\caption{Noise-induced pattern formation in simulations of phenotypic competition with a unit Gaussian competition kernel in (a) the weak-noise regime with mutation rate $D=10^{-3}$ and (b) the strong-noise regime with $D=10^{-5}$. Dark areas correspond to a high density of organisms and white space to empty regions of niche space. In both cases the macroscopic theory predicts that patterns should not form. This, and all other, simulations were conducting using the Gillespie algorithm \cite{Gillespie1977}. All simulations were initiated with a population of $K=10^3$ organisms distributed uniformly at random throughout niche space. To make the images, the locations of the organisms at times $t=0,1,2,\ldots$ were sorted into $1000$ bins, and a grayscale value assigned according to the bin occupancy.\label{Fig1}}
\end{figure}
\section{Stochastic model}
To initiate a proper treatment of demographic noise, we must start from a more fundamental description of the system, in terms of the interactions between individual organisms. We use the following microscopic dynamics: 
\begin{enumerate}
\item We model niche space as the one-dimensional line segment $(-\pi,\pi]$. For mathematical simplicity, and to eliminate end effects, we impose periodic boundary conditions.
\item At time $t$ there are $N(t)$ organisms with locations in niche space given by $x_1,\ldots,x_{N(t)}$.
\item Each organism reproduces with rate one. The position of the offspring is chosen from a normal distribution with variance $2D$, centred at the location of the parent.
\item The death rate of organism $i$ is $\frac{1}{K}\sum_jg(x_i-x_j)$, where, as above, $K$ controls the carrying capacity and $g$ is the competition kernel. We take $g$ to be positive, symmetric, and normalised so that $\frac{1}{2\pi}\int g(x) dx =1$.
\end{enumerate}
As we will show, equation \eqref{macro} can be derived directly from these dynamics in the limit of large population sizes. In that sense, the usual macroscopic theory is a coarse-grained description of this microscopic stochastic process. \par
A discrete-time version of this model was numerically simulated in \cite{Brigatti2008}, and found to exhibit stochastic patterns which depend on the shape of the kernel and other parameters. We will pursue a mathematical treatment of the problem with the advantages that (i) we will be able to make concrete, testable, predictions (ii) we can draw conclusions about the model on scales which are relevant to the real world, and hence too large to be simulated and (iii) most importantly, we gain deeper understanding of the mechanisms causing the phenomena we observe.\par
\section{Expansion of the master equation}
At a given moment in time, the state of the system is characterised by the distribution of individuals
\begin{equation}
\phi(x)=\frac{1}{K}\sum_{i=1}^{N(t)}\delta\left(x-x_i\right)\,.
\end{equation}
Our mathematical investigation of the model will focus on the time evolution of the probability $P(\phi,t)$ of finding the system in state $\phi$ at time $t$. To determine the rate of change of $P$ in time, we must consider contributions coming from the two processes which alter the system state -- birth and death.\par
The birth of an organism at location $y$ alters the distribution $\phi$ through the addition of another delta function. Similarly, the death of organism at $y$ corresponds to the removal of the delta peak there.  In finite-dimensional systems, it is usual to formalise creation and destruction processes using so-called step operators (see, for example, \cite{vanKampen1992}). We extend that formalism to our case by defining the operators $\Delta_y^+$ and $\Delta_y^-$, whose action on a generic functional $F[\phi(x)]$ is defined to be
\begin{equation}
\Delta^\pm_yF\Big[\phi(x)\Big]=F\Big[\phi(x)\pm\frac{1}{K}\delta(x-y)\Big]\,.
\end{equation}
The rate with which new organisms appear at point $y$ is the sum of the rates with which each existing organism reproduces there. Given that births happen with rate one and offspring appear at a Gaussian distance from their parent, we determine the birth rate functional to be
\begin{equation*}
\beta(x,\phi)=\sum_{i=1}^{N(t)} r(x-x_i)=K \int \phi(y)r(x-y)\,dy\,,
\end{equation*}
where we write $r(x)$ for the Gaussian probability density function with variance $2D$. By a similar logic, the total death rate $\gamma$ at point $x$ is determined by the product of the population density $\phi(x)$ with the death rate per organism there:
\begin{equation*}
\gamma(x,\phi)=\phi(x)\sum_{i=1}^{N(t)} g(x-x_i)=K \int \phi(x)\phi(y)g(x-y)\,dy\,.
\end{equation*}
Combining these two contributions, we obtain the functional master equation:
\begin{equation}
\begin{split}
\frac{\partial}{\partial t}P(\phi,t)=&K\int \bigg[\Big(\Delta^-_x-1 \Big)P(\phi,t)\beta(\phi,x)\\&\quad+\Big(\Delta^+_x-1 \Big)P(\phi,t)\gamma(\phi,x)\bigg]\,dx\,.
\end{split}
\label{MEQN}
\end{equation}
To make theoretical progress, we employ a functional variant of the Kramers-Moyal expansion \cite{vanKampen1992}, in which the $\Delta$ operators are re-written as Taylor series in powers $K^{-1}$:
\begin{equation}
\Delta_x^\pm=1\pm \frac{1}{K}\,\frac{\delta }{\delta\phi(x)}+\frac{1}{2K^2}\frac{\delta^2 }{\delta\phi(x)^2}\,+\mathcal{O}(K^{-3})\,,
\end{equation}
where $\delta/\delta\phi(x)$ denotes functional differentiation. Terminating the expansion at second order and applying to our master equation \eqref{MEQN} gives the functional Fokker-Planck equation
\begin{equation}
\begin{split}
\frac{\partial}{\partial t}P(\phi,t)&=-\int\int\frac{\delta }{\delta\phi(x)}\Big\{P(\phi,t)\mathcal{A}(\phi,x,y)\Big\}\,dx\,dy\\&+\frac{1}{2K}\int\int \frac{\delta^2 }{\delta\phi(x)^2}\Big\{P(\phi,t)\mathcal{B}(\phi,x,y)\Big\}\,dx\,dy\,,
\end{split}
\label{FFP}
\end{equation}
where 
\begin{equation}
\begin{split}
\mathcal{A}(\phi,x,y)&=\phi(y)\Big[r(x-y)-\phi(x)g(x-y)\Big]\,,\\
\mathcal{B}(\phi,x,y)&=\phi(y)\Big[r(x-y)+\phi(x)g(x-y)\Big]\,.
\end{split}
\end{equation}
This equation will form the basis of our mathematical treatment. Aside from its functional form, \eqref{FFP} differs from a generic Fokker-Planck equation in two regards: firstly, the non-local nature of the interactions mean that the drift term involves an integral over a second spatial variable; secondly, since the step operators in \eqref{MEQN} do not appear in pairs, the diffusion term is diagonal.\par
Depending on the scaling relationship between competition strength and mutation rate, the model exhibits two different types of behaviour, which we explore in turn.
\section{Weak-noise: stochastic patterns}
Sending $K\to\infty$ in \eqref{FFP} removes the diffusive part, leaving behind the Liouville equation corresponding to a deterministic variable $\phi(x,t)$ satisfying 
\begin{equation*}
\begin{split}
&\frac{\partial}{\partial t}\phi(x,t)=\int\phi(y,t)\Big(r(x-y)-\phi(x,t)g(x-y)\Big)\,dy\\
&\approx\phi(x,t)+D\frac{\partial^2}{\partial x^2}\phi(x,t)-\int \phi(x,t)\phi(y,t)g(x-y)\,dy\,,\\
\end{split}
\end{equation*}
where the second line comes from assuming $D$ is sufficiently small that only the linear term contributes. Identifying $\Phi(x,t)=K\phi(x,t)$, we obtain precisely the macroscopic model \eqref{macro}.\par
The central limit theorem suggests that the state of the microscopic model should be well described by the above deterministic equation, plus some stochastic fluctuations of order $K^{-1/2}$. We look for such fluctuations around the homogeneous state $\phi(x)\equiv1/2\pi$ by changing variables to
\begin{equation}
\zeta(x)=\sqrt{K}\left(\phi(x)-\frac{1}{2\pi}\right)\,,
\end{equation}
whence \eqref{FFP} becomes, to leading order in $K$,
\begin{equation*}
\begin{split}
\frac{\partial}{\partial t}P(\zeta,t)=&-\int\int A(x-y)\frac{\delta }{\delta\zeta(x)}\Big\{\zeta(y)P(\zeta,t)\Big\}\,dx\,dy\\
&+\int \frac{\delta^2 }{\delta\zeta(x)^2}\,P(\zeta,t)\,dx\,,
\end{split}
\end{equation*}
where $A(x)=r(x)-g(x)-\delta(x)$. This equation can be solved in Fourier space: writing $\zeta_n$ and $A_n$ for the modes of $\zeta$ and $A$, we find that the $\zeta_n$ are independent (up to the condition $\zeta_{-n}=\overline{\zeta_n}$ ), zero mean complex Gaussian random variables with variances $\langle|\zeta_n|^2\rangle$ evolving according to 
\begin{equation}
\frac{d}{dt}\big\langle|\zeta_n|^2\big\rangle=2A_n\big\langle|\zeta_n|^2\big\rangle+\frac{1}{\pi}\,.
\end{equation}
Here, and hereafter, angle-brackets refer to the average over the noise. Sending $t\to\infty$ here and changing variables back to $\phi(x)$ leads to the following description of the long-time behaviour of the population density:
\begin{equation}
\phi(x)= \frac{1}{2\pi}+\frac{1}{\sqrt{K}}\sum_{n=-\infty}^\infty \zeta_n e^{inx}\,.
\label{PZ}
\end{equation}
To compare our theoretical prediction with the results of simulations, it is useful to define a statistic to characterise the structure of the typical states of the system. We choose the following scaled measure of spatial covariance:
\begin{equation}
\Xi(x)=2\pi\int\Big\langle \phi(y)\phi(y-x)\Big\rangle_\infty \,dy\,,
\label{defXi}
\end{equation}
where $\langle\cdots\rangle_\infty$ denotes averaging with respect to the stationary (long-time) distribution of $\phi$. This quantity is particularly useful as it tells us if the organisms form localised clusters or not: a completely flat spatial covariance $\Xi(x)\equiv 1$ would indicate a homogeneous distribution of organisms, while a sharp peak at the origin reveals the presence of species. In the weak-noise regime, equation \eqref{PZ} leads to the expression $\Xi(x)=1+\frac{1}{K}\sum_n\langle|\zeta_n|^2\rangle e^{inx}$. The agreement between simulations and theory is excellent: see Figure 2 for an example.\par
\begin{figure}
\includegraphics[width=0.42\textwidth, trim=0 20 60 0]{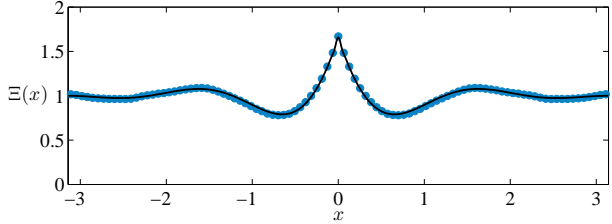}
\caption{Spatial covariance in the weak-noise regime, with parameters corresponding to those of Figure 1(a). Blue circles show the empirical result approximated by 100 bins and averaged over 100 independent simulation runs, while the black line is the theoretical prediction of equation \eqref{PZ}. The peak at $x=0$ indicates the formation of clusters.}
\end{figure}
Many complex systems, particularly those relating to biological and chemical processes, exhibit stochastic pattern formation in space and/or time \cite{McKane2005,Butler2008}. This effect can be thought of as the stochastic shadow of patterns which occur elsewhere in the parameter space of the corresponding deterministic system, and it is particularly strong when parameters are chosen so that the deterministic system is close to the transition between patterned and pattern-free phases. This is also true in our case, where the period of the stochastic patterns corresponds to the Fourier mode which is nearest to becoming unstable. \par
As evidenced in Figures 1 and 2, for moderate values of $K$ the stochastic patterns can be quite pronounced, making the typical states of the system appear very different to the homogeneous distribution predicted by equation \eqref{macro}.

\section{Strong-noise: spontaneous species formation} Our results in the weak noise regime are based on treating the mutation rate $D$ as fixed while taking the limit of large carrying capacity $K$. We will now see that when this assumption breaks down, the behaviour of the model is radically different. In what follows, we will study the special case of a flat competition function ($g(x)\equiv1$), meaning that all organisms experience the same death rate and there is no selective pressure at all\footnotemark[1]. Unsurprisingly, the macroscopic theory predicts that no species should form in this case, that is, the homogeneous distribution is globally stable.\par\footnotetext[1]{We should point out that, as well as being the most important case from a paradigmatic point of view, the flat competition kernel is a prerequisite for the mathematical techniques we apply.}
The mathematical study of demographic noise in this regime is considerably less straightforward; one can no longer appeal to the standard techniques that work for weak noise, and must instead develop an approach specific to the problem. Our method is to observe that the population as a whole, ignoring phenotypic differences, undergoes logistic growth and should thus rapidly converge to a value close to the carrying capacity $K$. This fact may be exploited mathematically through the adiabatic elimination of the zeroth Fourier mode of the population density.\par
As before, we take equation \eqref{FFP} for the starting point of our analysis, this time expressed in Fourier space. If $\phi_n$ are the modes of $\phi$, then their joint distribution obeys the complex Fokker-Planck equation
\begin{equation}
\begin{split}
&\frac{\partial P}{\partial t}=-\sum_{n} \frac{\partial}{\partial\phi_n}\Big\{\phi_n\big(e^{-Dn^2}-2\pi\phi_0\big)P\Big\}\\&+\frac{1}{4K\pi}\,\sum_{n,m} \frac{\partial}{\partial \phi_n}\frac{\partial}{\partial\phi_{m}}\Big\{\phi_{n+m}\big(e^{-D(n+m)^2}-2\pi\phi_0\big)P\Big\}\,.
\end{split}
\label{NLFP}
\end{equation}
We are able to isolate $\phi_0$ from the rest of the system by integrating over the other modes to obtain an equation describing noisy logistic growth:
\begin{equation}
\begin{split}
\frac{\partial P}{\partial t}=&-\frac{\partial}{\partial\phi_0}\Big\{\phi_0(1-2\pi\phi_0)P\Big\}\\&+\frac{1}{4K\pi}\,\frac{\partial^2}{\partial \phi_0^2}\Big\{\phi_{0}(1+2\pi\phi_0)P\Big\}\,.
\end{split}
\end{equation}
Ignoring the absorbing state at zero (corresponding to extinction), the quasi-stationary distribution of $\phi_0$ may be found by means of a Wentzel-Kramers-Brillouin expansion \cite{Risken1989}. Briefly, the procedure is as follows: adoption of the ansatz $P(\phi_0)=\exp\{-K\,S(\phi_0)\}$ reduces the above Fokker-Planck equation to a Hamilton-Jacobi equation for $S(\phi_0)$, plus some correction terms of smaller order in $K^{-1}$. Solving the Hamilton-Jacobi equation and transforming back gives the approximate solution
\begin{equation}
P(\phi_0)\propto e^{-4 K\pi\phi_0}(1+2\pi\phi_0)^{4K}\,.
\end{equation}
For large $K$, this distribution approaches a peak of width approximately $(4K\pi^2)^{-1}$, centred at $1/2\pi$.\par
The programme now is to assume that $\phi_0$ rapidly approaches a value close to $1/2\pi$, and to approximate the behaviour of the other modes based on this premise. The most intuitive approach is to work with a stochastic differential equation, rather than the Fokker-Planck equation. Choosing the $\text{It}\bar{\text{o}}$ formalism, equation \eqref{NLFP} (for the flat kernel case) is equivalent to the Langevin system
\begin{equation}
\frac{d}{dt}\phi_n=\phi_n\big(e^{-Dn^2}-2\pi\phi_0\big) + \frac{1}{\sqrt{2\pi K}}\,\eta_n(t)\,,
\end{equation}
where the $\eta_n(t)$ are zero mean complex Gaussian white noises with covariance structure
\begin{equation}
\Big\langle \overline{\eta_n(t)}\,\eta_m(t^\prime)\Big\rangle =\delta(t-t^\prime)\phi_{n+m}\big(e^{-D(n+m)^2}+2\pi\phi_0\big)\,.
\end{equation}
The modes must also jointly satisfy the requirement that they define the Fourier series of a non-negative function. Fortunately, we are able to evaluate the statistics of their amplitude without dealing explicitly with this constraint. Inserting the approximation $\phi_0(t)\equiv 1/2\pi$ into the $n=0$ equation yields
\begin{equation}
\eta_0(t)\equiv0\,,
\label{zeronoise}
\end{equation}
and for the other equations
\begin{equation}
\frac{d}{dt}\phi_n=\phi_n\big(e^{-Dn^2}-1\big) + \frac{1}{\sqrt{2\pi K}}\,\eta^\prime_n(t)\,,
\label{langevin}
\end{equation}
where the covariance structure of the $\eta^\prime_n(t)$ is that of the original system, conditioned upon the event \eqref{zeronoise}. Specifically,
\begin{equation*}
\begin{split}
\Big\langle \overline{\eta^\prime_n(t)}\,\eta^\prime_m(t^\prime)\Big\rangle=&\delta(t-t^\prime)\bigg\{\phi_{n+m}\big(e^{-D(n+m)^2}+1\big)\\
&-\pi\phi_{n}\phi_{m}\big(e^{-Dn^2}+1\big)\big(e^{-Dm^2}+1\big)\bigg\}\,.
\end{split}
\end{equation*}
A differential equation for the moments $\big\langle|\phi_n|^2\big\rangle$ can now be obtained directly from \eqref{langevin}, for example by $\text{It}\bar{\text{o}}$'s formula \cite{Ito1951}. Specifically, we find
\begin{equation*}
\begin{split}
&\frac{d}{dt}\big\langle|\phi_n|^2\big\rangle=\\
&\quad\left[2\big(e^{-Dn^2}-1\big)-\frac{1}{2K}\big(e^{-Dn^2}+1\big)^2\right]\big\langle|\phi_n|^2\big\rangle+\frac{1}{2K\pi^2}\,.
\end{split}
\end{equation*}
We are now in a position to examine the behaviour of the system under the joint scaling $K\to\infty$ and $D\to 0$, where the product of carrying capacity and mutation rate approaches a fixed value; we choose $KD\to\tau$. In this limit we obtain a compact expression for the spatial covariance:
\begin{equation}
\Xi(x)=\sum_{n=-\infty}^\infty\frac{1}{1+ n^2\tau }\,e^{inx}\,.
\label{XI}
\end{equation}
This result demonstrates that, in the strong-noise regime, the macroscopic equation \eqref{macro} is not a correct description of the system as the effects of demographic noise cause completely different behaviour. Figure 3 shows an example of the spontaneous emergence of a phenotypic cluster, along with comparison between simulations and equation \eqref{XI} for the same parameter values.\par
\begin{figure}
\includegraphics[width=0.43\textwidth, trim=0 10 60 15]{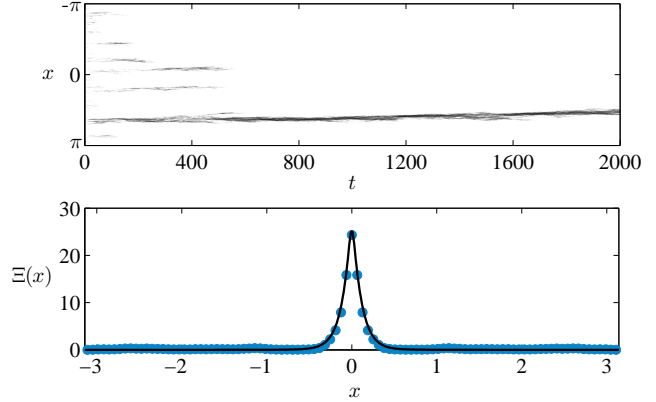}
\caption{Spontaneous species formation with a flat competition kernel and $KD=0.01$ ($K=10^3$, $D=10^{-5}$). Blue circles show the empirical result approximated by 100 bins and averaged over 100 independent simulation runs, while the black line is the theoretical prediction of equation \eqref{XI}. The strong peak at $x=0$ indicates the formation of species.}
\end{figure}
The shape of $\Xi(x)$ for $\tau$ ranging over several orders of magnitude is shown in Figure 4. When $\tau$ is very small, the spatial covariance is sharply peaked at the origin, meaning that the organisms form a single very tight phenotypic cluster. As we move to larger values of $\tau$, the typical width of this cluster increases; the species is becoming more phenotypically diverse. Eventually as $\tau\to\infty$ we move through the weak-noise regime to a limiting case where no species form and the spatial covariance is completely flat. Although the calculation presented here applies to the case of a flat competition kernel, the effect of spontaneous speciation is robust to changes in the kernel. For non-flat kernels it is typical for several distinct species to emerge, see Figure 1(b) for an example.
\begin{figure}
\includegraphics[width=0.4\textwidth, trim=0 17 50 -4]{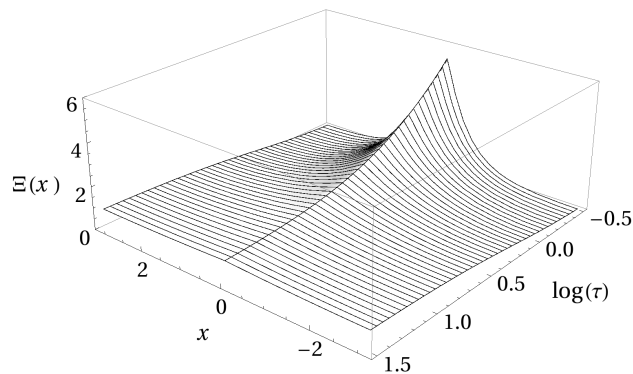}
\caption{Limiting spatial covariance in the strong-noise regime for a range of $\tau$ values. }
\end{figure}
\section{Discussion}
Simple mathematical models of the type inspired by MacArthur and Levins \cite{MacArthur1967} provide a useful framework in which the effects of phenotypic competition can be studied. Theoretical work on this topic has in the past been limited to the macroscopic level, described by equations such as \eqref{macro}. The implicit assumption of such work is that the effects of demographic noise are not important and may be ignored. In the present analysis we have taken a ground-up approach to the problem, beginning with a description of the system on the fundamental level of interactions between individual organisms and culminating in a thorough treatment of the effects of demographic noise. In doing so, we have uncovered behaviour which is radically different from that predicted by the macroscopic theory. \par
Depending on the scaling of mutation rate $D$ and carrying capacity $K$, two different types of noise effects are present. If $D$ is held constant in the limit $K\to\infty$ the typical states of the system can be expressed in terms of fluctuations around the homogeneous fixed point of the macroscopic equations; we refer to this as the weak-noise regime. The precise relationship is given by equation \eqref{PZ} and, as evidenced in Figures 1(a) and 2, the typical states of the system are very different to the homogeneous distribution predicted by the macroscopic theory. In the strong-noise regime, where the mutation rate and carrying capacity scale together as $KD\to \tau$, we find that demographic noise can become strong enough to cause the spontaneous formation of well-defined stable species. We are able to make theoretical progress by studying this phenomenon in the paradigmatic case of the flat competition kernel, where the limiting form \eqref{XI} of the spatial covariance can be computed. Figure 4 provides an overview of the different possible behaviours, interpolating between a homogeneous distribution and the formation of tightly localised species. Together, these results demonstrate that consideration of demographic noise completely overhauls our understanding of these models. \par
The fact that models of phenotypic competition are capable of very different behaviours depending on the relationship between mutation rate and carrying capacity may have important implications for our understanding of speciation in the natural world. Across different taxa, typical population sizes vary by many orders of magnitude, whilst it is known that there is considerably less variation in the rate of genetic mutation \cite{Makarieva2004}. Interpreted in the context of the present work, our theoretical results suggest that organisms belonging to taxa with very large global populations (bacteria, foraminifers, {\it etc}) may form only very loose ecological species which could be difficult to tell apart. By contrast, the comparatively smaller population sizes common to larger creatures imply the spontaneous formation of well-defined species. This observation may help explain the apparent difficulty in precisely identifying species of meiofauna \cite{Creer2010}, an hypothesis that will receive a careful examination in future work. \par

The biological interpretation of our results in the context of species formation is, of course, rather broad since the model does not include sexual reproduction, inhomogeneity in niche space or any number of other possible complicating factors. Whilst limiting the realism of the model, it can be argued that this reductionist approach actually strengthens our conclusions; the other factors are certainly important in understanding how and where species form, but our work shows they cannot be solely responsible, since species will form spontaneously even in their absence. 
\par
In a wider context, our results highlight the importance of theoretical approaches to phenomena of this type. With current computing power, individual-based simulations of complex systems involving a realistic number of participants are not usually possible. It is common to postulate that the results of a small simulation can be treated as representative of the world at large. However, we have shown here that as the number individual agents in a simulation increases, radically different behaviours are possible depending on how demographic noise interacts with the other aspects of a model. This exemplifies the power of theoretical methods to provide useful and interesting results that could not be obtained from simulations.

\begin{acknowledgments}
TR acknowledges funding from the EPSRC under grant number EP/H02171X/1. AGR was supported by a Beaufort Marine Research Award by the Marine Institute, under the Sea Change strategy and the Strategy for Science, Technology and Innovation, funded under the Irish National Development Plan (2007-2013).
\end{acknowledgments}

\end{document}